# MontiArcAutomaton: Modeling Architecture and Behavior of Robotic Systems

Jan Oliver Ringert*, Bernhard Rumpe, and Andreas Wortmann

*Abstract*—Robotics poses a challenge for software engineering as the vast numbers of different robot platforms impose different requirements on robot control architectures. The platform dependent development of robotic applications impedes reusability and portability. The lack of reusability hampers broad propagation of robotics applications.

The MontiArcAutomaton architecture and behavior modeling framework provides an integrated, platform independent structure and behavior modeling language with an extensible code generation framework. MontiArcAutomaton's central concept is encapsulation and decomposition known from Component & Connector Architecture Description Languages. This concept is extended from the modeling language to the code generation and target runtime framework to bridge the gap of platform specific and independent implementations along well designed interfaces. This facilitates the reuse of robot applications and makes their development more efficient.

## I. INTRODUCTION

Robotics poses a challenge for software engineering. There are vast numbers of robots ranging from industrial manipulators, over service robots, to autonomous cars and space explorers. Most of these robots define their own unique hard- and software architectures. The software for these robots is usually implemented by robotics domain experts using general-purpose programming languages [1]. This lack of abstraction leads to monolithic solutions for limited problems [2] which hampers reuse and prevents the development of new robotics applications [3].

Within the last decade reuse for robotics software has intensively been pursued. Most of this research focuses on applying some form of component-based based software development to robotics [4], [5], [6], [7], [8], [9]. These platform-specific building blocks encapsulate domain knowledge, but their development and integration still require robotics experts to be programming experts.

Model-based development of software architecture and behavior of robotics applications with abstraction from the target platforms helps domain experts to focus on their tasks. The technical implementation details are handled by code generators. While existing modeling tools for robotics architectures [1], [10] focus on architecture and communication issues, we have developed the integrated architecture and behavior modeling framework MontiArcAutomaton [11], [12], [13] for robotic systems. The goals of the MontiArcAutomaton framework are:

The authors are with the Software Engineering department, RWTH Aachen University, Germany, http://www.se-rwth.de.
*J.O. Ringert is supported by the DFG GK/1298 AlgoSyn.

1) target platform independent development of robotic control software,
2) problem specific modeling of control behavior,
3) support for reuse and portability of components and libraries, and
4) efficient deployment to different target platforms with support for native implementations.

We address the first two goals with the MontiArcAutomaton modeling language and the latter two with the code generation and library concepts.

## II. MONTIARCAUTOMATON MODELING

The language MontiArcAutomaton inherits from the Architecture Description Language MontiArc [14] to model distributed and hierarchically decomposed robotics systems as Component & Connector architectures. Components are either atomic and their behavior is directly defined, or they are composed from other components and their behavior is derived from the composition. Communication between components is only possible via unidirectional connectors between the typed ports of components. We distinguish between the definition and the instantiation of components. MontiArc allows multiple instantiation of components and supports the definition of generic components, which can be instantiated for different types, and configurable components, which can be configured with concrete values at their instantiation. MontiArc thus imposes stable component interfaces as identified necessary [1], [5], [10] and therefore promotes a separation of concerns into system integration and component implementation.

Figure 1 illustrates these concepts on the architecture of a simple robot called `BumperBot` which traverses an area by driving straight until hitting an obstacle, it then drives backwards a little, rotates and drives straight forward again. The robot consists of five components, where the `controller` receives input from a touch sensor via the port `bump` and sends movement commands to two motors via the ports `left` and `right`. To determine the time driving backwards, the `BumperBot` additionally contains a `Timer` which is set by `controller` using the port `timer`. The component implementations of `TouchSensor` and `Motor` are platform specific. While the models themselves remain stable, their implementations are imported from a corresponding platform specific library.

Component behavior in MontiArcAutomaton can be modeled platform independently using the I/O$^\omega$ automata [15], [16] paradigm. The behavior of the atomic component

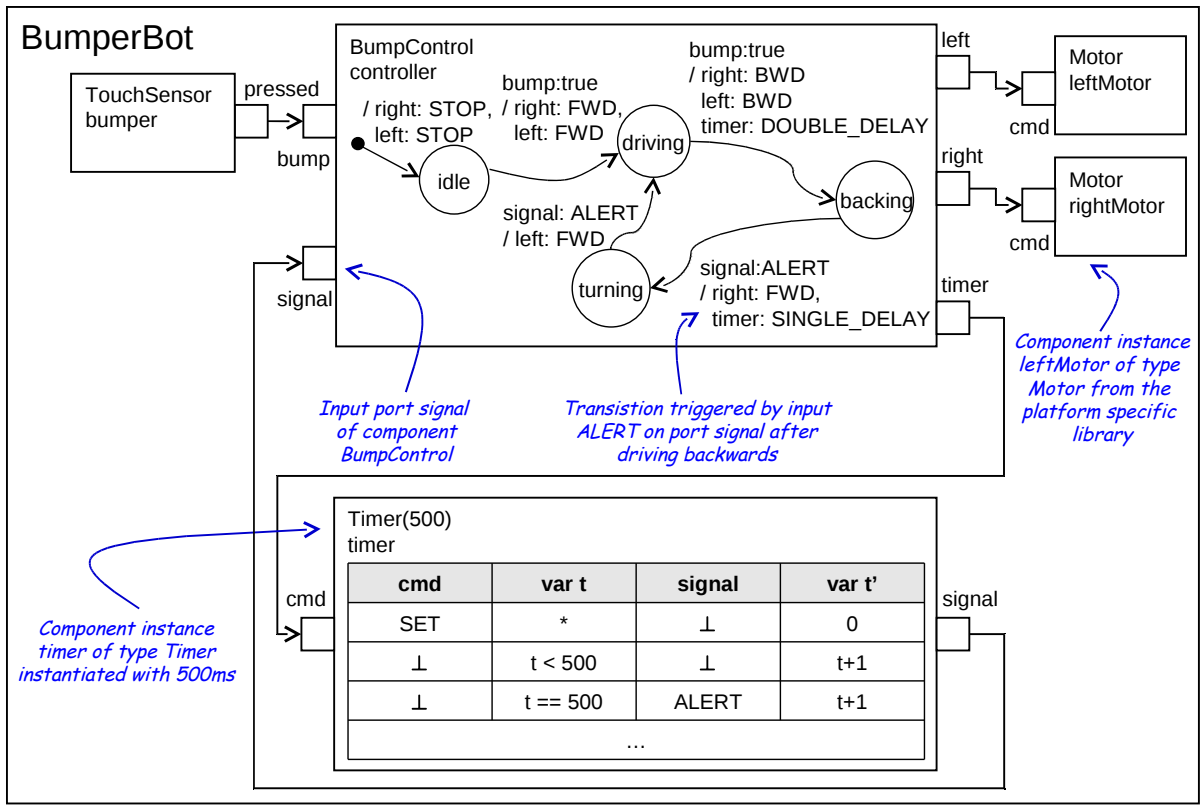

Fig. 1. Software architecture and behavior model of the `BumperBot` robot. The behavior of component `controller` is implemented as an I/O$^\omega$ automaton. The behavior of component `timer` is implemented as a set of rules.

`controller`, which defines how the robot reacts to an obstacle, is modeled as an I/O$^\omega$ automaton over the ports of the embedding component. We are currently experimenting with embedding further domain specific languages into MontiArc component definitions. As an example, the behavior of component `timer` is implemented using a language of rules over ports and variables of the component. The first line describes that after receiving the command SET on port `cmd` independent of the current value of variable `t`, the component emits no signal on port `signal` (denoted by $\bot$) and `t` is set to 0.

## III. CODE GENERATION AND PLATFORM-SPECIFIC CODE

MontiArcAutomaton provides code generators for several target languages [12]. Currently supported target languages are Mona (for formal analysis), EMF Ecore (for graphical editing), Java, and Python (both for deployment). The code generators for Java and Python additionally support the robot operating system (ROS) [7] either by means of rosjava[1] or natively. The framework further provides an Eclipse[2] based editor for the textual and graphical development of MontiArcAutomaton models.

We have developed a model library of standard components that need target and runtime specific implementations. MontiArcAutomaton code generators support the generation of wrapper components that can be configured with a specific implementation without modifying the generated code. This mechanism allows to provide hand written implementations that persist regeneration of code from models. We use the same mechanism for all generated code, which enables us, e.g., for testing purposes, to replace the manual or generated implementations of components.

Currently, MontiArcAutomaton only supports component composition, automata, and native code implementations of components. We plan to extend the support for other modeling languages, e.g., the rules language illustrated in component `timer` in Fig. 1, through MontiCore's language embedding mechanisms [17]. This way engineers will be able to implement behavior using automata, native code, and their favorite DSLs while the composition of components to systems is – oblivious to their implementation – on the architecture level using MontiArc with explicitly defined component interfaces.

## IV. CONCLUSION

We have presented the MontiArcAutomaton framework and modeling language, which allow platform independent modeling of robot control architectures and behavior as Component & Connector architectures. The models provide stable interfaces and can be decomposed to be developed independently. MontiArcAutomaton models of the software architecture are platform independent and can be transformed to platform specific implementations by generators for different target platforms. These can easily be developed using the compositional MontiArcAutomaton code generation framework. The MontiArcAutomaton framework is based on MontiCore and thus extensible with different behavior modeling languages up to implementations in native programming languages.

---

[1] The rosjava website: https://code.google.com/p/rosjava/.
[2] The Eclipse foundation website: http://www.eclipse.org/.

## REFERENCES

[1] C. Schlegel, T. Hassler, A. Lotz, and A. Steck, "Robotic software systems: From code-driven to model-driven designs," in *Advanced Robotics, 2009. ICAR 2009. International Conference on*, 2009, pp. 1–8.

[2] P. J. Mosterman, "Elements of a Robotics Research Roadmap: A Model-Based Design Perspective," in *Workshop on Manufacturing and Automation*, Washington, D.C., 2008.

[3] M. Hägele, N. Blümlein, and O. Kleine, "Wirtschaftlichkeitsanalysen neuartiger Servicerobotik- Anwendungen und ihre Bedeutung für die Robotik-Entwicklung," BMBF, Tech. Rep., 2011. [Online]. Available: http://www.ipa.fraunhofer.de/

[4] A. Brooks, T. Kaupp, A. Makarenko, S. Williams, and A. Oreback, "Towards component-based robotics," in *Intelligent Robots and Systems, 2005.(IROS 2005). 2005 IEEE/RSJ International Conference on*. IEEE, 2005, pp. 163–168.

[5] D. Brugali and P. Salvaneschi, "Stable Aspects In Robot Software Development," *International Journal of Advanced Robotic Systems*, vol. 3, 2006.

[6] D. Brugali, A. Brooks, A. Cowley, C. Côté, A. Domínguez-Brito, D. Létourneau, F. Michaud, and C. Schlegel, "Trends in Component-Based Robotics," in *Software Engineering for Experimental Robotics*, ser. Springer Tracts in Advanced Robotics, D. Brugali, Ed. Berlin, Heidelberg: Springer Berlin Heidelberg, 2007, vol. 30, ch. 8, pp. 135–142.

[7] M. Quigley, B. Gerkey, K. Conley, J. Faust, T. Foote, J. Leibs, E. Berger, R. Wheeler, and A. Ng, "ROS: an open-source Robot Operating System," in *ICRA Workshop on Open Source Software*, 2009.

[8] T. Niemueller, A. Ferrein, D. Beck, and G. Lakemeyer, "Design Principles of the Component-Based Robot Software Framework Fawkes," in *Proc. of Second International Conference on Simulation, Modeling, and Programming for Autonomous Robots*, ser. Lecture Notes in Computer Science. Darmstadt, Germany: Springer, 2010.

[9] S. Thierfelder, V. Seib, D. Lang, M. Häselich, J. Pellenz, and D. Paulus, "Robbie: A Message-based Robot Architecture for Autonomous Mobile Systems," in *INFORMATIK 2011 - Informatik schafft Communities*, H.-U. Heiß, P. Pepper, H. Schlingloff, and J. Schneider, Eds. Köllen Druck+Verlag GmbH Bonn., 2011.

[10] C. Schlegel, A. Steck, and A. Lotz, "Model-Driven Software Development in Robotics : Communication Patterns as Key for a Robotics Component Model," in *Introduction to Modern Robotics*, D. Chugo and S. Yokota, Eds. iConcept Press, 2011.

[11] J. O. Ringert, B. Rumpe, and A. Wortmann, "A Requirements Modeling Language for the Component Behavior of Cyber Physical Robotics Systems," in *Modelling and Quality in Requirements Engineering*. Monsenstein und Vannerdat Münster, 2012, pp. 133–146.

[12] J. O. Ringert and B. Rumpe and A. Wortmann, "From Software Architecture Structure and Behavior Modeling to Implementations of Cyber-Physical Systems," in *Software Engineering 2013 Workshopband*, ser. LNI, Stefan Wagner and Horst Lichter, Ed., vol. 215. GI, Köllen Druck+Verlag GmbH, Bonn, 2013, pp. 155–170.

[13] "MontiArcAutomaton project web page," http://www.se-rwth.de/materials/ioomega/, 2013, Accessed 3/13.

[14] A. Haber, J. O. Ringert, and B. Rumpe, "Montiarc - architectural modeling of interactive distributed and cyber-physical systems," RWTH Aachen, Tech. Rep. AIB-2012-03, february 2012.

[15] B. Rumpe, *Formale Methodik des Entwurfs verteilter objektorientierter Systeme*. Herbert Utz Verlag Wissenschaft, 1996.

[16] J. O. Ringert and B. Rumpe, "A Little Synopsis on Streams, Stream Processing Functions, and State-Based Stream Processing," *International Journal of Software and Informatics*, vol. 5, no. 1-2, pp. 29–53, July 2011.

[17] H. Krahn, B. Rumpe, and S. Völkel, "MontiCore: a framework for compositional development of domain specific languages," *STTT*, vol. 12, no. 5, pp. 353–372, 2010.